\documentclass{aa}

\usepackage{graphicx}

\def\M{{\cal M}}
\def\rco{{r_{\rm co}}}
\def\cs{c_{\rm s}}
\def\ie{i.e. }
\def\eg{e.g. }
\def\e{{\rm e}}
\def\d{{\rm d}}
\def\beq{\begin{eqnarray}}
\def\eeq{\end{eqnarray}}
\def\v{\upsilon}
\def\vr{\upsilon_r}
\def\vphi{\upsilon_\varphi}
\def\om{\omega}
\def\omp{\omega'}
\def\p{\partial}
\def\rsh{r_{\rm sh}}
\def\rs{r_{\rm son}}
\def\fsh{f_{\rm sh}}
\def\gsh{g_{\rm sh}}
\def\Bsh{B_{\rm sh}}
\def\drho{\delta\rho}

\def\dvr{\delta\upsilon_r}
\def\dvphi{\delta\upsilon_\varphi}
\def\dwz{\delta w_z}
\def\lam{\lambda}
\def\na{\nabla}

\def\dxi{\Delta\xi}
\def\T{\tau}

\begin{document}

\title{Non-axisymmetric instabilities in shocked accretion flows with
differential rotation}

\author{Wei-Min Gu\inst{1,2}
          \and
          Thierry Foglizzo\inst{1}
         }

\offprints{guwm@xmu.edu.cn}

\institute{Service d'Astrophysique, CEA/DSM/DAPNIA, CEA-Saclay,
               91191 Gif-sur-Yvette, France
            \and
           Department of Physics, Xiamen University,
               Xiamen, Fujian 361005, People's Republic of China
            }

\date{}

\titlerunning{Non-axisymmetric instabilities in shocked accretion flows}
\authorrunning{Gu \& Foglizzo}

\abstract{
The linear stability of a shocked isothermal accretion flow
onto a black hole is investigated in the inviscid limit.
The outer shock solution, which was previously found to be stable with
respect to axisymmetric perturbations, is, however, generally unstable to
non-axisymmetric ones.
Eigenmodes and growth rates are obtained by numerical integration of
the linearized equations.
The mechanism of this instability is based on the cycle of acoustic
waves between their corotation radius and the shock.
It is a form of
the Papaloizou-Pringle instability, modified by
advection and the presence of the shock. As such it can be
generalized to non isothermal shocked accretion flows. Blobs and
vortices are generated by the shock as a by-product of the
instability.
}

\maketitle

\section{Introduction}

Hydrodynamic instabilities
of shocked accretion flows may explain some of the properties of X-ray 
binaries, such as their time variability. The structure of stationary 
accretion flows involving shocks was described by Fukue (1987) and 
Chakarabarti (1989a,b). Even with the simple inviscid hypothesis, 
the structure of shocked accretion flows is complex, and their stability 
is not yet fully understood. As noted by Nakayama (1992), the calculations 
of Chakrabarti (1989a,b) is a study of the forced oscillations of the flow 
rather than an analysis of its intrinsic stability. Nakayama (1992, 1993) 
introduced a new type of global instability
between the inner sonic point and the shock. 
He found that, of the two possible shock positions, the inner one is
unstable due to post-shock acceleration, while the outer one is stable
due to post-shock deceleration. His conclusion was confirmed
by Nobuta and Hanawa (1994), whose numerical simulations showed that the
inner shock is completely destroyed by perturbations, while
the outer one is stable. All the above works, however, were based on
axisymmetric calculations.
Molteni, T\'oth \& Kuznetsov (1999, hereafter MTK) performed 2-D simulations of
an adiabatic flow with an outer shock and found a non-axisymmetric
instability.
They showed that the instability saturates at a low
level, and a new asymmetric
configuration develops, with a deformed shock rotating steadily.
MTK pointed out that this effect may have relevant observational
consequences, such as quasi-periodic oscillations (QPO).
The mechanism of the instability was not explained by MTK,
who briefly mentioned a possible link with the numerical simulations
of Blaes \& Hawley (1988). The Papaloizou-Pringle instability (1984,
hereafter PPI)
simulated by Blaes \& Hawley is known to take place in
discs or tori,
in which the radial velocity is initially zero,
whereas the flow simulated by MTK involves radial advection,
an inner sonic point and an outer shock.
The effect of advection on the PPI was investigated by Blaes (1987),
who found that the PPI is strongly
stabilized by advection at the inner boundary. The interpretation
of the results of MTK in terms of the PPI is thus not obvious {\it a
priori}.
What is more, the flow studied by MTK is also potentially
unstable
by the advective-acoustic mechanism  (Foglizzo \& Tagger 2000,
Foglizzo 2001, 2002),
based on the cycle of entropy/vorticity
perturbations and acoustic
waves in the subsonic region between a
stationary shock and a sonic surface. \\
The aim of this study is thus to understand the instability
mechanism at work in shocked accretion flows.
For the sake of
simplicity, the present linear stability analysis is
focused on isothermal accretion flows with constant angular momentum.
The paper is organised as follows. Linearized equations and boundary
conditions are described in Section 2 and solved numerically in
Section 3. The instability mechanism is analysed in Section 4, and
compared to the simulations by MTK in Section 5.

\section{Equations}

\begin{figure}
\centering
\includegraphics[width=8cm]{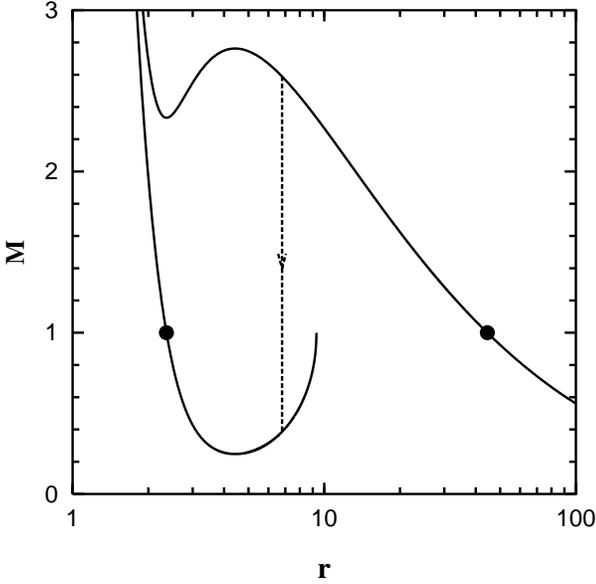}
\caption{Radial profile of the Mach number in the unperturbed flow,
for which $l=1.87$ and $\cs=0.1$. The two circles denote
the inner and the outer sonic points, respectively. The dashed arrow
shows the shock at $\rsh = 6.8$.
}
\label{figprofile}
\end{figure}
An inviscid, isothermal flow around a black hole is considered,
in the pseudo-Newtonian potential introduced by Paczy\'nski and
Wiita (1980), $\Phi \equiv -GM/(r-r_{\rm g})$.
Equations are made
dimensionless by using the Schwarzschild radius
and the speed of light as reference units, i.e., $r_{\rm g} \equiv 1$
and $c \equiv 1$. In these units, the Keplerian frequency is
noted
$\Omega_{\rm K}=1/(r-1)(2r)^{1\over2}$.
In order to overcome the
technical difficulty of treating a realistic vertical structure,
three types of simplifying assumptions can be used; constant
thickness, constant angle (conical flows), or vertical equilibrium.
Chakrabarti \& Das (2001) proved that
there is no essential difference among these three assumptions from
the point of view of the existence of stationary solutions.
The thickness of the flow is approximated as a
constant for the sake of
simplicity, as in Nakayama (1992), Nobuta \& Hanawa (1994),
Blaes (1987) or MTK, although the approximate thickness $H\sim
\cs/\Omega_{\rm K}$ deduced from the balance of the vertical
pressure force and the vertical gravity is not constant.

The stationary flow is described by the conservation of mass and the
Bernoulli equation:
\beq
\rho r v_r&=& {\rm const}\ ,\label{mass}
\eeq
\beq
{v_r^2\over 2}+{l^2\over 2r^2}+\cs^2\log{\rho}-{1\over 2(r-1)}&=&{\rm const}\ ,
\label{bernoulli}
\eeq
where $\rho$ is the density, $\vr$ is the radial velocity,
$\cs$ is the sound speed, and $l$ is the specific angular momentum.
An example solution of unperturbed flow is shown in Fig.~\ref{figprofile},
for which $l=1.87$ and $\cs=0.1$.
As a consequence of the inviscid hypothesis, stationary accretion 
flows onto a black hole are very sub-Keplerian in their outer parts. 
The question of the origin of this sub-Keplerian flow is still uncertain 
(Chakrabarti \& Titarchuk 1995, Molteni et al. 2001).  
In Fig.~\ref{figprofile}, the unperturbed flow
is supersonic between the outer sonic point ($R_{\rm s2} = 44.5$) and
the shock ($\rsh = 6.8$), becomes subsonic
between the shock and the inner sonic point ($R_{\rm
s} = 2.36$) and goes into
the central black hole supersonically . This example solution is
precisely the one chosen by
Nobuta and Hanawa (1994), in which they showed that the outer shock is stable
to axisymmetric perturbations. Our numerical results in Section~3, 
however, indicate that it is unstable to non-axisymmetric perturbations.

The mass conservation equation and the Euler equation are
written as follows:
\beq
\frac{\p\rho}{\p t}+\na \cdot (\rho\v) = 0 \ ,
\label{masscon}
\eeq
\beq
\frac{\p\v}{\p t}+w\times \v +\na \left\lbrack\frac{\v ^2}{2}
+\cs^2\log \rho -\frac{1}{2(r-1)}\right\rbrack = 0 \ ,
\label{Euler}
\eeq
where $w$ is the vorticity.
In order to write the linearized equations in the simplest form, the
two functions $f,g$
are defined as follows:
\beq
f &\equiv& \frac{\drho}{\rho}+\frac{\vr \dvr}{\cs^2}
+\frac{\vphi \dvphi}{\cs^2} \ ,
\label{feq}
\eeq
\beq
g &\equiv& \frac{\drho}{\rho}+\frac{\dvr}{\vr} \ ,
\label{geq}
\eeq
where $f$ is the perturbation of the Bernoulli constant and $g$ is the
perturbation of the mass accretion rate.
The frequency $\omp$ of the perturbation measured in the rotating
frame is defined as:
\beq
\omp \equiv \om - m\Omega \ ,
\eeq
where $m$ is the azimuthal wave number and $\Omega \equiv l/r^2$ is
the angular velocity.
With the standard method of linear stability analysis, (perturbations
proportional to $e^{-i\om t+im\varphi}$, see Appendix B), the
following differential system is obtained:
\beq
\frac{\p f}{\p r} &= & \frac{i\om \M^2}{\vr (1-\M^2)}
\left(-\frac{\omp}{\om}f+g\right)  \nonumber\\
&& +\frac{l\Bsh}{r^2\cs^2\vr (1-\M^2)}
\e^{\int_{\rsh}^r\frac{i\omp}{\vr}\d r} \ ,\label{dfdr}\\
\frac{\p g}{\p r} &= & \frac{i\omp}{\vr (1-\M^2)}
\left(\frac{\omp}{\om}f-\M^2g\right)
- \frac{im^2\cs^2}{\om r^2\vr}f   \nonumber\\
&& - \frac{\Bsh}{\om r^2\vr}\left\lbrack m+\frac{\omp l}{\cs^2(1-\M^2)}
\right\rbrack \e^{\int_{\rsh}^r\frac{i\omp}{\vr}\d r} \ ,\label{dgdr}
\eeq
where $B \equiv r\vr w_z$, $w_z$ is the vorticity along the rotation axis,
$\M\equiv -\vr /\cs$ is the radial Mach number, and the
subscript 'sh' denotes the shock position.
The boundary conditions corresponding to a perturbed shock velocity
$\Delta v_r$ are obtained in Appendix C:
\beq
\fsh &=& \left(
\frac{\omp}{\om}-\frac{i\M\cs\eta}{\om r}
-{1\over1-\M^2}\right)
(1-\M^2)^2\frac{\Delta \vr}{\vr} \ ,\label{fsh}\\
\gsh &=& \frac{\omp}{\om}(1-\M^2)\frac{\Delta \vr}{\vr} \ ,\label{gsh}\\
\Bsh &=& -im\cs^2\left(\frac{\omp}{\om}-\frac{i\M\cs\eta}{\om r}\right)
(1-\M^2)^2\frac{\Delta \vr}{\vr} \ ,\label{Bsh}
\eeq
where $\eta \equiv (d\log \M / d\log r)|_{\rm sh}$ (Eq.~\ref{defeta}),
$\M$ and $\vr$ are calculated at the post-shock side. \\
The 
vorticity $\dwz$ produced by the shock is:
\beq
\dwz = \frac{\Bsh}{r\vr}\e^{\int_{\rsh}^r\frac{i\omp}{\vr}\d r} \ .\label{wz}
\eeq
There are two differential equations for $f,g$ and one unknown parameter
$\om$ in our system, thus three boundary conditions are needed to solve
the equations. In addition to the two boundary conditions
Eqs.(\ref{fsh}-\ref{gsh})
at the shock, a third equation is obtained from the critical
condition at the sonic point,
\beq
\om g_{\rm son}-\omp f_{\rm son}-\frac{il\Bsh}{\rs^2\cs^2}
\e^{\int_{\rsh}^{\rs}\frac{i\omp}{\vr}\d r}= 0 \ .\label{critic}
\eeq
These three boundary conditions are used to numerically solve the
differential system Eqs.~(\ref{dfdr}-\ref{dgdr}) and to determine the
eigenfrequencies $\om$.
A single equation corresponding to this boundary value problem is
formulated in Appendix~D.

\section{Numerical results}

\begin{figure}
\centering
\includegraphics[width=8cm]{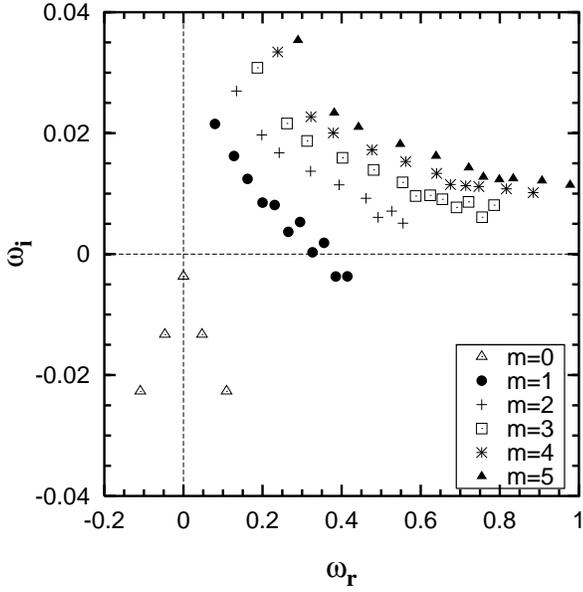}
\caption{
Eigenspectrum of the flow $l=1.87$ and $\cs=0.1$, showing the instability
of the modes $1\le m\le 5$. The empty triangles correspond to the stable
axisymmetric eigenmodes found by Nobuta and Hanawa (1994).
}
\label{figNH94}
\end{figure}
\begin{figure}
\centering
\includegraphics[width=8cm]{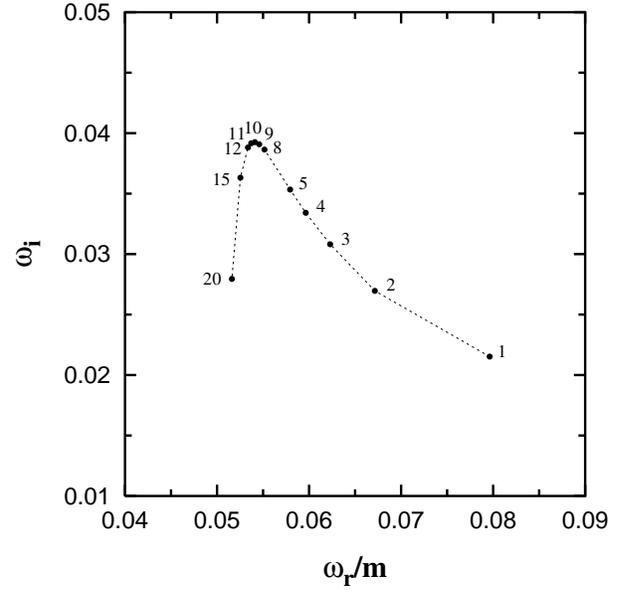}
\caption{
The maximum growth rate of the non radial modes  is reached for $m\sim 10$
in  the flow $l=1.87$ and $\cs=0.1$. The pattern frequency $\omega_r/m$ of
these unstable modes is concentrated in a limited band.
}
\label{figmax10}
\end{figure}
The standard Runge-Kutta method is used to integrate differential equations
from the sonic point to the shock. Fig.~\ref{figNH94} shows the eigenspectrum
of the flow $l=1.87$ and $\cs=0.1$ studied by Nobuta \& Hanawa (1994),
for perturbations $0\le m \le 5$. The stability of axisymmetric perturbations
confirms the results of Nobuta \& Hanawa (1994).
In contrast, non-axisymmetric perturbations are unstable. The highest
growth rate is reached for $m=10$ perturbations, as seen in
Fig.~\ref{figmax10}. Determining numerically how the
most unstable $m$ depends on the value of $(l,\cs)$ is beyond the scope
of this paper.
In what follows, numerical calculations are restricted to $m = 1$
perturbations for the sake of simplicity.\\
\begin{figure}
\centering
\includegraphics[width=8cm]{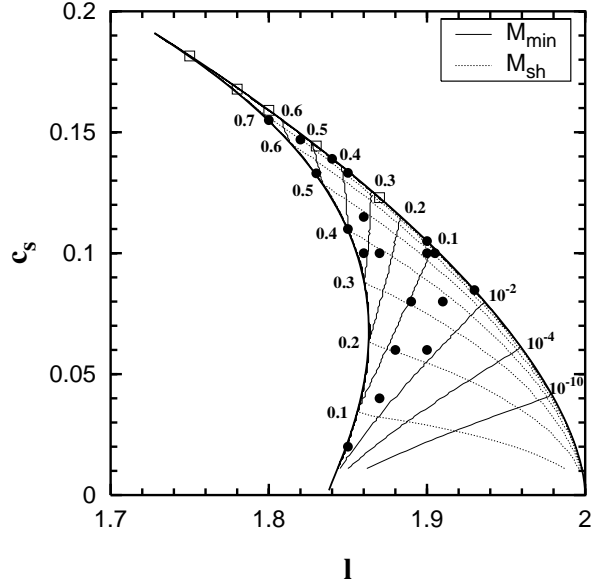}
\caption{
The two thick solid lines are the threshold for shock-included solutions.
The dotted lines measure the shock strength by the value of $\M_{\rm
sh}$
indicated on the left.
The thin solid lines correspond to the
value of the minimum
Mach number $\M_{\rm min}$ indicated on the
right.
The 19 filled circles correspond to unstable outer shock solutions, and the
5 empty squares correspond to stable ones.
}
\label{figlcs}
\end{figure}
Fig.~\ref{figlcs} shows the domain $(l,\cs)$ of angular momentum and sound
speed for which an isothermal inviscid flow, subsonic far from the
accretor, may accrete onto a black-hole through a stationary shock.
This domain is limited by two solid lines.
The solid line on the
left corresponds to the
solutions in which the inner shock and the outer shock are identical,
whereas the solid line on the right corresponds to
extremely weak shocks, $\M_{\rm sh}\sim 1$.
A total of 24 shocked solutions were considered for numerical calculation,
in order to get a broad view of their stability properties.
Since the inner shock solution was already found
unstable to axisymmetric perturbations, we concentrated on the stability
of the outer shock solution. Solutions at the bottom right corner
in Fig.~\ref{figlcs}, for which the advection timescale is much longer
than the acoustic one ($\M_{\rm min} \le 0.01$),
were avoided for computing time reasons.
Among this sample of 24 shocked flows, 19 were found to be unstable and 5
to be stable. The results shown in Fig.~\ref{figlcs} suggest that the
shock is generally unstable to non-axisymmetric perturbations, except
for a very narrow region close to the upper right border. In other words,
the shock might be stable only for $\M_{\rm sh}\sim 1$.

\section{Instability mechanism}

\subsection{Comparison with the advective timescale}

\begin{figure}
\centering
\includegraphics[width=8cm]{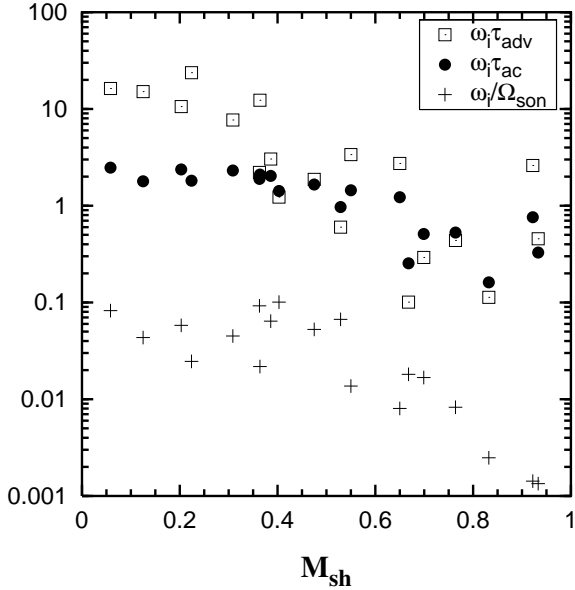}
\caption{Growth rates compared to three timescales: advection
$\tau_{\rm adv}$ (squares), acoustic $\tau_{\rm ac}$ (circles), and
rotation $\Omega(\rs)$ (crosses).
}
\label{figwscale}
\end{figure}
In an isothermal flow, two types of cycles may exist between the sonic
point and the shock, i.e., the purely acoustic cycle and the vortical-acoustic
cycle. From the results of Foglizzo (2002), we would expect
the vortical-acoustic cycle to be particularly unstable for very strong shocks,
$\M_{\rm sh}\ll 1$, with a growth time comparable to the advection timescale
$\T_{\rm adv}$:
\beq
\T_{\rm adv} \equiv \int_{\rs}^{\rsh}\frac{\d r}{|\vr|} \ .
\eeq
The numerical results seem to exclude an explanation based on
a vortical-acoustic cycle:
\par (i) the flow is generally unstable to non-axisymmetric perturbations,
even for mild shocks: the range of shock strengths among the $19$
unstable flows considered is very wide ($0.06\le\M_{\rm sh}\le0.94$).
\par(ii) the growth time can be much shorter than the advection time.
The growth time of the instability is plotted in Fig.~\ref{figwscale},
in units of the rotation frequency at the sonic radius $\Omega(\rs)$,
the advection time $\T_{\rm adv}$ and also in units of
the acoustic time $\T_{\rm ac}$ approximated by the following lower bound:
\beq
\T_{\rm ac} \equiv {2\over1-\M_{\rm min}^2}{\rsh-\rs\over\cs} \ .
\eeq
The dispersion of the points in Fig.~\ref{figwscale} is smallest
when the growth
rate is measured in units of the acoustic time,
suggesting a purely
acoustic cycle.
If this is the case, what is the mechanism ?

\subsection{The corotation region in the PPI mechanism}

\begin{figure}
\centering
\includegraphics[width=8cm]{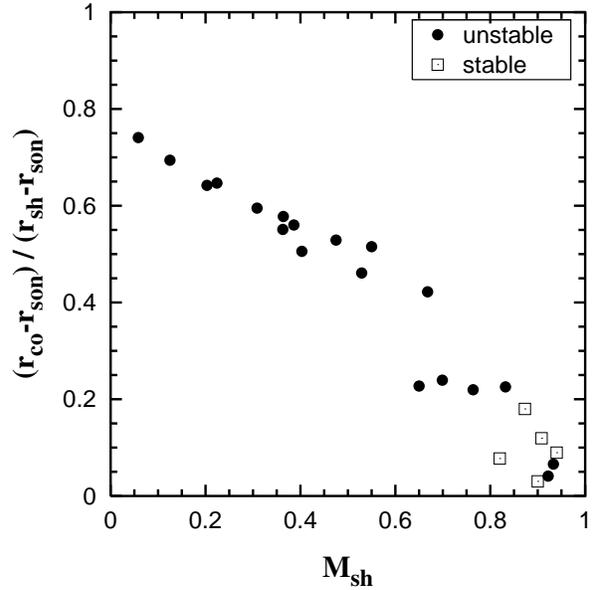}
\caption{
The position of the corotation radius is indicated for 19 unstable
flows (circles) and 5 stable ones (squares).
}
\label{figcorot}
\end{figure}
\begin{figure}
\centering
\includegraphics[width=8cm]{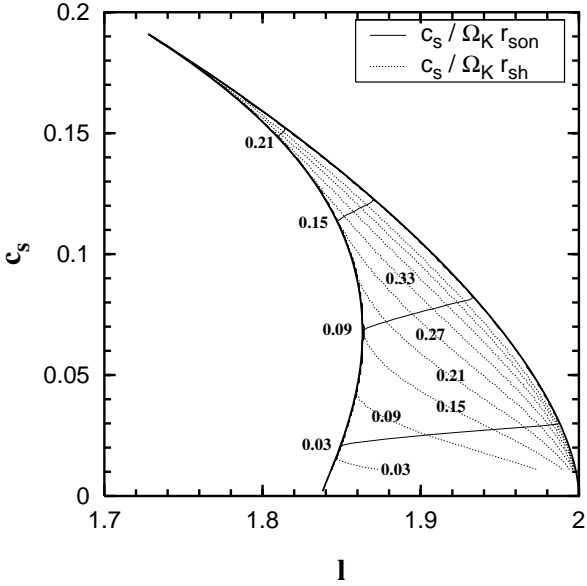}
\caption{
The value of the ratio $\cs/\Omega_{\rm K}r$ at the sonic point
(dotted lines) and at the shock (full lines) is indicated on the
lines.
}
\label{figH}
\end{figure}
The corotation radius $\rco$ of the perturbation is defined by
$\omega= m\Omega_0$, where $\Omega_0\equiv \Omega(\rco)$:
\beq
\rco \equiv \left({lm\over\om
_r}\right)^{1\over2} \ .
\eeq
According to Fig.~\ref{figcorot}, the
corotation radius of the most unstable modes
is always located between the sonic point and the 
shock.
Fig.~\ref{figwscale} and Fig.~\ref{figcorot}
are hints in favour of the PPI.
The mechanism of the PPI was
formulated most simply by Goldreich \& Narayan 1985 (hereafter
GN85) in thin discs. The more subtle effect of a corotation
resonnance (Narayan, Goldreich \& Goodman 1987, Kato 1987) can be
neglected in the present study, since the epicyclic frequency is
zero ($\Omega=l/r^2$). The PPI is based on the exchange of energy
and angular momentum between acoustic waves propagating inside and
those propagating outside the corotation radius (Mark 1976).
The thin disc hypothesis in these studies is technically important in
order to treat high frequency perturbations in the WKB approximation
$\omega=m\Omega_0\gg \cs/r$.  According to Fig.~\ref{figH}, the high
frequency approximation is applicable only in the limit $\cs\to 0$,
which coincides with strong shocks according to Fig.~\ref{figlcs}.
Following GN85 and Kato (1987), we believe that
the physical mechanism captured analytically in thin discs is also
relevant in thicker discs ($\Omega_{\rm K} r/c_s\sim 1$), even if a
quantitative estimate of the growth rate is precluded.

The PPI
occurs with either an inner or an outer reflecting boundary, or even
more efficiently with both. Let us recall this mechanism in the case
of a single reflecting boundary, noting ${\cal T}_{\rm co}$ the
fractional amplitude of the transmitted wave through the corotation
zone, and ${\cal R}$ the fractional amplitude of the acoustic wave
reflected by the boundary.  An incident wave with energy $+1$
transmits an energy $-|{\cal T}_{\rm co}|^2$ on the other side of
corotation, thus amplifying the energy of the reflected wave by a
factor  $1+|{\cal T}_{\rm co}|^2$. If $\tau$ is the duration of the
acoustic cycle, the growth rate $\omega_i$ of this instability
is:
\beq
\omega_i={1\over\tau}\log\left\lbrack|{\cal R}|(1+|{\cal
T}_{\rm co}|^2)^{1\over 2}\right\rbrack \ .\label{omegi}
\eeq
The
following estimate of ${\cal T}_{\rm co}$ was obtained by GN85 and
confirmed by Kato (1987):
\beq
{\cal T}_{\rm
co}\sim\exp\left\lbrack-{\pi\over 2} {\cs\over
m|{\dot\Omega}_0|}
\left({m^2\over\rco^2}+{\kappa^2\over\cs^2}\right)\right\rbrack \ .
\label{tco}
\eeq
The
instability is thus slow in thin discs if $\kappa>0$, and the most
unstable modes correspond to $m\sim \kappa \rco/\cs$. Let us now
investigate the effects of radial advection on this
mechanism.

\subsection{Effect of advection on the corotation region}

The effect of advection on the corotation region can be
handled analytically in the high frequency limit.
The second order
differential equation satisfied by the homogeneous solution $f_0$ is
obtained in Appendix~E:
\beq
\left\lbrace{\p^2 \over\p
r^2}+k^2\right\rbrace
\left\lbrack\left({1-\M^2\over\M}\right)^{1\over2}f_0
\e^{-\int^r
i\omp{\M\over1-\M^2}{\d r\over
\cs}}\right\rbrack=0 \ ,\label{difcanon}
\eeq
In the high frequency
limit, acoustic waves can propagate in the region where the function
$k^2(r)$ is positive, and are evanescent in the region where
$k^2(r)<0$. $k^2$ is approximated at high frequency as
follows:
\beq
k^2&\sim&
{1\over(1-\M^2)}
\left\lbrack{\omp^2\over(1-\M^2)\cs^2}-{m^2\over
r^2}\right\rbrack \ .\label{k2}
\eeq
$k^2$ is positive near the sonic
point ($\M\sim1$), negative near the corotation radius
($\omp\sim0$),
and positive again at large radius. A calculation with
$\kappa=0$ and $\M>0$ in Appendix~F
shows that ${\cal T}_{\rm co}$ is
independent of the Mach number in the high frequency approximation,
if the gradient of the Mach number is neglected:
\beq
{\cal T}_{\rm
co}\sim\exp\left\lbrack-{\pi\over 2} {m\cs\over
|{\dot\Omega}_0|\rco^2}\right\rbrack \ ,\label{tcom}
\eeq
where
${\dot\Omega}_0$ is the derivative of the rotation frequency at the
corotation radius.
This estimate at high frequency thus favours modes
with a low azimuthal number $m$. Beyond the high frequency
approximation, this calculation shows that the amplification
mechanism at corotation can operate even with radial
advection.

\subsection{Crucial role of the boundaries}

\begin{figure}
\centering
\includegraphics[width=8cm]{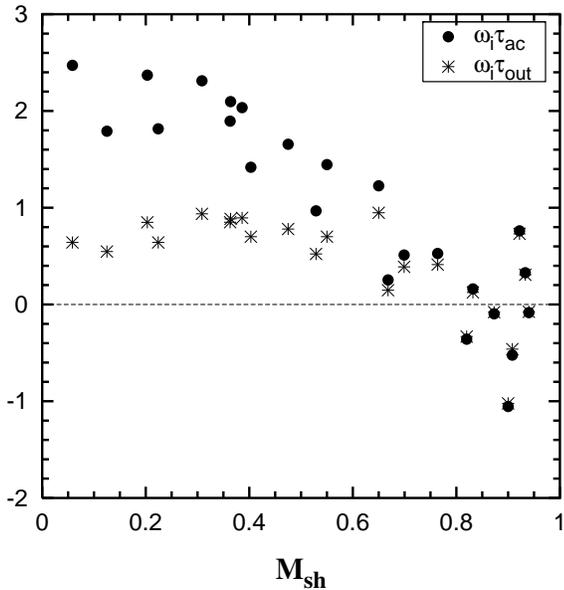}
\caption{
Growth rate measured in units of the total acoustic time $\tau_{\rm
ac}$ (circles)
compared to the acoustic time $\tau_{\rm out}$
between the corotation and the shock (stars). }
\label{figwitac}
\end{figure}
In order to estimate the effect of radial advection at
the inner boundary, Blaes (1987) considered
an inviscid flow with
constant thickness in a pseudo-Newtonian
potential, and a uniform angular momentum. It differed from the
present study by
an adiabatic hypothesis and, most importantly, a
leaking outer boundary.
As stressed by GN85, the PPI requires at
least one reflecting
boundary. The stabilizing effect of advection found by Blaes (1987)
is thus directly
related to the impossibility, for acoustic waves
moving away from the corotation,
to be reflected towards it.
By contrast, a stationary shock standing at the outer boundary of the flow
is a good reflector of acoustic waves, if the shock is not too weak.
A continuity argument would lead to expect the reflection coefficient
${\cal R}_{\rm sh}$ to decrease to zero in the limit $\M_{\rm sh}\to
1$. This is confirmed by a calculation of ${\cal R}_{\rm sh}$ in the
high frequency limit (see Appendix~E), which gives the same result as
in isothermal flows without rotation (Foglizzo 2002):
\beq
{\cal R}_{\rm sh}\sim- {1-\M_{\rm sh}\over1+\M_{\rm sh}} \ .
\label{rshwkb}
\eeq
According to Eq.~(\ref{omegi}) and
(\ref{rshwkb}), a strong isothermal shock (\ie $|{\cal R}_{\rm
sh}|\sim 1$) is a sufficient condition for the instability of the
acoustic cycle between the shock and corotation in thin discs.
Acoustic energy in shocked flows is thus trapped between the shock
and the
corotation radius, and may leak inside through the sonic radius without
damping the instability.
On the basis of the results obtained by Blaes (1987), a strong decrease of
the growth rate should be expected in the weak shock limit.
This
trend is clear in Fig.~\ref{figwscale}, when measuring the growth
rate,
as Blaes (1987), in units of the rotation frequency at a fixed
radius.
What is more, all the stable modes correspond to weak shocks
($\M_{\rm sh}>0.8$) according to Fig.~\ref{figcorot}.\\
Fig.~\ref{figwitac} shows that the dispersion of the points is
further decreased
by measuring the growth rate in units of the
acoustic timescale $\tau_{\rm out}$ between
the corotation and the
shock, approximated by:
\beq
\tau_{\rm out}\sim {2\over1-\M_{\rm
min}^2}{r_{\rm sh}-\rco\over \cs} \ .\label{tauout}
\eeq
\begin{figure}
\centering
\includegraphics[width=8cm]{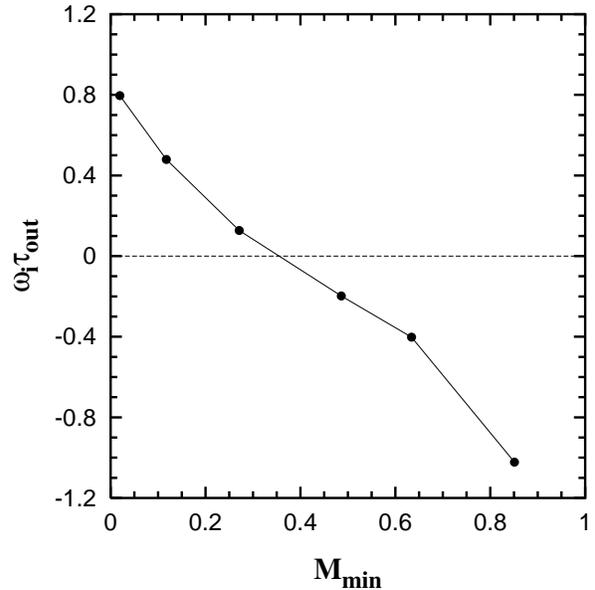}
\caption{Incidence of advection on the PPI for weak shocks ($\M_{\rm
sh}=0.9$). The growth rate of the instability is measured in units of
the acoustic time $\tau_{\rm out}$ between the corotation and the
shock.}
\label{figMmini}
\end{figure}
The combination of Eqs.~(\ref{omegi}), (\ref{tcom}),
(\ref{rshwkb}) and (\ref{tauout}) gives an analytical estimate of the
growth rate which is justified only in the high frequency limit, \ie
for $\cs\to 0$ and strong shocks. A direct application of this
estimate outside its range of validity is unable to explain
the strong dispersion of the points in Fig.~\ref{figwitac} for weak
shocks.
As noted in Sect.~4.2 with Fig.~\ref{figH}, solutions with a
weak shock coincides with those with the largest parameter
($\cs/\Omega_{\rm K}r$), \ie the least adapted to a high frequency
approximation.
It should be noted that part of the dispersion of the
growth rate in Fig.~\ref{figwitac} may come from the contribution of
the vorticity perturbations, when the advection time is comparable to
the acoustic time: they trigger additionnal acoustic waves which may
add or substract to the acoustic power present in the acoustic cycle.
Could this effect be essentially stabilizing ? According to
Fig.~\ref{figMmini}, for a given shock strength (\ie $\M_{\rm
sh}=0.9$), the flow seems to be stabilized by a strong advection.

\section{Comparison with adiabatic simulations}

Although
non-linear processes might redistribute the energy of the unstable
modes,
it could be tempting to identify the frequency $\omega _r^{\rm
sim}$, observed in the
simulations, with the real part of the most unstable eigenmode.
Based on five regular oscillation examples, MTK
found that the oscillation period $P$ is always in the same range,
$60 < P < 280$ despite the large variation in the rotation rate at
the shock distance, $100 < 2\pi /\Omega _{\rm sh} < 2000$.
We find that all the five regular oscillation examples listed in MTK satisfy
the condition,
$\Omega_{\rm sh} < \omega _r^{\rm sim} \equiv 2\pi /P < \Omega_{\rm son}$,
which is a requirement of the PPI mechanism.\\
A quantitative
comparison with the results of MTK is hampered by our
choice of an
isothermal flow rather than an adiabatic one, and also by our
restriction
to $m=1$ perturbations which are not necessarily the most
unstable ones.
A global understanding of which azimuthal wavenumber
corresponds to the
most unstable perturbation has not been reached yet.
The only isothermal example considered in Fig.~3 with $m>1$
perturbations
favours $8\le m\le12$, which seems higher than the low
azimuthal wavenumbers
present in the simulations of MTK. \\
Despite these uncertainties, the present isothermal calculation
is simple
enough to illustrate the possibility of a PPI in a shocked accretion
flow, occuring on a shorter time scale than any possible
advective-acoustic cycle.
This instability mechanism should thus also apply to adiabatic flows,
including those simulated by MTK, if the shock is able to reflect
acoustic waves. The calculation of ${\cal R}_{\rm sh}$ in the
adiabatic hypothesis (Foglizzo \& Tagger 2000, Eq.~(C13)) would
be unchanged by rotation and non radial
perturbations, in the high frequency limit, for a strong shock:
\beq
{\cal R}_{\rm sh}\sim
-{2^{1\over2}-\gamma^{1\over2}(\gamma-1)^{1\over2}
\over
2^{1\over2}+\gamma^{1\over2}(\gamma-1)^{1\over2}} \ .\label{rshadiab}
\eeq
Comparing Eq.~(\ref{rshadiab}) with Eq.~(\ref{rshwkb}),
the reflection
coefficient of strong adiabatic shocks is significantly smaller than
strong isothermal ones
in the high frequency limit.
The adiabatic
hypothesis in the work of MTK may thus marginally influence the
strength of the instability ($|{\cal R}_{\rm sh}|\sim 0.36$ for
$\gamma=4/3$).\\
A refined comparison of the growth rate and
oscillation frequency with the results of MTK would require
a linear
stability analysis in an adiabatic flow involving also $m\ge 2$
perturbations.
This will be the subject of a forthcoming paper.

\section{Conclusion}

The main results can be summarized as follows: \\
1. Despite post-shock deceleration, the outer shock solution is generally
unstable with respect to non-axisymmetric perturbations. \\
2. The growth time is comparable to the acoustic time if the shock
is not too weak. The flow might be stable only when $\M_{\rm sh}$ is 
close to $1$. \\
3. A WKB analysis of thin discs with advection 
proves that the acoustic cycle between 
the corotation radius and the 
shock is unstable if the shock is strong, thus extending the 
mechanism of the Papaloizou-Pringle instability to shocked discs 
with advection.\\
4. Extrapolating the results established in thin 
discs, the instability of thicker discs could be interpreted as a 
form of the Papaloizou-Pringle instability modified by advection. 
This interpretation is supported by the existence of a corotation 
radius within the disc, a growth time comparable to the acoustic 
time, and the role of the shock strength. The influence of advection 
in the specific case of weak shocks, however, is still unexplained 
(Fig.~\ref{figMmini}).\\

One particularity of this instability 
should be pointed out:
although the instability mechanism relies on acoustic waves,
vortical waves are continuously generated by the perturbed shock
and advected to the accretor (Eqs.~\ref{Bsh}, \ref{wz}). These vortical waves
(or the entropy/vorticity waves in the case of
non isothermal flows) may play the role of the 'blobs' invoked
frequently to model the time variability of X-ray binaries
(\eg references in Belloni et al. 2002). The observed properties of 
these blobs could therefore be used to test the validity of a model 
of shocked accretion.

\begin{acknowledgements}
The authors would like to thank Dr M. Tagger for discussions. W. Gu
is grateful to Prof. Lu for his encouragements.
\end{acknowledgements}

\appendix

\section{Description of the unperturbed flow}

We consider a simple one-dimensional steady flow under the assumption
of a constant thickness. The derivative of the Mach number is
obtained
by derivating the stationary flow equations (\ref{mass}-\ref{bernoulli}):
\beq
\eta\equiv \frac{d\log \M}{d\log r} = -\frac{1+(l^2-l_{\rm K}^2)/r^2\cs^2}
{1-\M^2} \ ,\label{defeta}
\eeq
where $l_{\rm K}$ is the Keplerian angular momentum,
$l_{\rm K}^2 \equiv r^3/2(r-1)^2$.
The continuity of the flow at the sonic point requires
\beq
\rs^2\cs^2+l^2-\frac{\rs^3}{2(\rs-1)^2} = 0 \ .
\eeq
The $(l,\cs)$ parameter space for the shocked accretion flows is shown
in Fig.~\ref{figlcs}.

\section{Linearized equations for perturbations}

The mass conservation and Euler equations
are linearized in order to
obtain the following three equations:
\beq
\vr \frac{\p g}{\p r}+\frac{im}{r}\dvphi -i\omp \frac{\drho}{\rho} = 0 \ ,\\
\cs^2\frac{\p f}{\p r}-i\om \dvr -\frac{l}{r}\dwz = 0 \ ,\\
\frac{im\cs^2}{r}f-i\om \dvphi +\vr \dwz = 0 \ .\label{vphiwz}
\eeq
The evolution of vorticity is deduced from the curl of Euler equation:
\beq
\frac{\p w}{\p t}+w(\na \cdot v)+(v\cdot \na )w-(w\cdot \na )v = 0 \ .
\eeq
The vorticity can be integrated when linearized, leading to Eq.~(\ref{wz}).
The differential system of Eqs.(\ref{dfdr}-\ref{dgdr}) is obtained from
the above equations.
The perturbation $\dvphi$ is related to $f$ and $\Bsh$
(Eqs.~(\ref{vphiwz},\ref{wz})),
\beq
\delta\vphi = \frac{1}{i\om r}\left(im\cs^2f
+\Bsh \e^{\int_{\rsh}^r\frac{i\omp}{\vr}\d r}\right) \ .
\label{dvphi}
\eeq
Eqs.~(\ref{dfdr}-\ref{dgdr}), (\ref{dvphi}) enable us to get
$(\dvr ,\dvphi ,\drho)$ from $(f,g)$.

\section{Boundary conditions at the shock}

The isothermal shock conditions are written as follows:
\beq
\rho_-\v_{r-} = \rho_+\v_{r+} \ ,
\eeq
\beq
\rho_-(\v_{r-}^2+\cs^2) = \rho_+(\v_{r+}^2+\cs^2) \ ,
\eeq
\beq
\v_{\varphi -} = \v_{\varphi +} \ ,
\eeq
where the subscripts "-" and "+" denote pre-shock and post-shock quantities,
respectively. Let the shock be perturbed by $\dxi$ in the radial
direction:
\beq
\dxi \propto \e^{-i\om t+im\varphi} \ .
\eeq
The perturbed velocity $\Delta \vr$ of the shock and its angle
$\Delta\theta$ with the azimuthal direction are
\beq
\Delta \vr &=& \frac{\p \dxi}{\p t} = -i\om \dxi \ ,
\eeq
\beq
\Delta\theta &=& \frac{1}{r}\frac{\p \dxi}{\p \varphi}
= \frac{im}{r}\dxi = -\frac{m}{\om r}\Delta \vr \ .
\eeq
Considering the velocity component perpendicular to the shock, in the frame
of the shock, the isothermal shock conditions become the following:
\beq
\left(\rho_- + \frac{\p \rho_-}{\p r}\dxi\right)
\left(\v_{r-}+\frac{\p \v_{r-}}{\p r}\dxi -\frac{\omp}{\om}\Delta \vr\right)
= \nonumber\\
\left(\rho_+ + \frac{\p \rho_+}{\p r}\dxi +\drho\right)\nonumber\\
\times
\left(\v_{r+}+\frac{\p \v_{r+}}{\p r}\dxi -\frac{\omp}{\om}\Delta \vr
+\dvr\right) \ ,\\
   \left(\rho_- + \frac{\p \rho_-}{\p r}\dxi\right)
\left\lbrack\left(\v_{r-}+\frac{\p \v_{r-}}{\p r}\dxi -\frac{\omp}{\om}
\Delta \vr\right)^2
+\cs^2\right\rbrack  \nonumber\\
= \left(\rho_+ + \frac{\p \rho_+}{\p r}\dxi +\drho\right)  \nonumber\\
\times
\left\lbrack\left(\v_{r+}+\frac{\p \v_{r+}}{\p r}\dxi -\frac{\omp}{\om}
\Delta \vr +\dvr\right)^2
+\cs^2\right\rbrack \ ,\\
\v_{\varphi -}+\frac{\p \v_{\varphi -}}{\p r}\dxi +\v_{r-}\Delta\theta=
\nonumber\\
\v_{\varphi +}+\frac{\p \v_{\varphi +}}{\p r}\dxi +\v_{r+}\Delta\theta
+\delta\vphi \ .
\eeq
The boundary conditions Eqs.~(\ref{fsh}-\ref{Bsh}) are deduced from the
above three equations.

\section{Boundary value problem}

With three boundary conditions
Eqs.(\ref{fsh}-\ref{gsh}),(\ref{critic}), the differential
system
could in principle be
directly integrated using the Runge-Kutta method.
The singularity at the sonic point requires to integrate from the sonic
point towards the shock. The boundary value problem is conveniently
reduced to a
single equation involving the solution $f_0$ of the homogeneous system.
The homogeneous differential equations are expressed as follows:
\beq
\frac{\p f_0}{\p r} &=& \frac{i\om \M^2}{\vr (1-\M^2)}
\left(-\frac{\omp}{\om}f_0 + g_0\right) \ ,\label{df0}\\
\frac{\p g_0}{\p r} &=& \frac{i\omp}{\vr (1-\M^2)}
\left(\frac{\omp}{\om}f_0-\M^2g_0\right)
- \frac{im^2\cs^2}{\om r^2\vr}f_0 \ .\label{dg0}
\eeq
At the sonic point, the following boundary conditions are obtained:
\beq
g_0|_{\rm son} =\frac{ \omp}{\om}f_0|_{\rm son} \ .\label{gson}
\eeq
The derivation at the sonic point can be obtained from L'Hospital method,
\beq
\frac{\p f_0}{\p r}|_{\rm son} &=& \frac{1}{2\rs}\frac{\frac{\omp^2\rs^2}
{\cs^2}-m^2
-\frac{2ilm}{\cs\rs}}{\eta_{\rm son}+\frac{i\omp\rs}{\cs}} f_0|_{\rm son}\ ,
\label{dfson}\\
\frac{\p g_0}{\p r}|_{\rm son} &=& \left(\frac{im^2\cs}{\om r^2}-\frac{i\omp^2}
{\om \cs}\right)f_0|_{\rm son}
-\frac{\omp}{\om}\frac{\p f_0}{\p r}|_{\rm son}\ ,\label{dgson}\\
\eta_{\rm son}&=& -{1\over \cs}
\left\lbrack{\rs(\rs+1)\over 4(\rs -1)^3}-\left({l\over \rs}\right)^2\right
\rbrack^{1\over2}.\label{etason}
\eeq
With the method in 'Appendix C' of Foglizzo (2002),
the following single equation for the boundary value problem is obtained:
\beq
\left(f\cdot\frac{\p f_0}{\p r}-\frac{\p f}{\p r}\cdot f_0\right)|_{\rm sh}
\nonumber \\
= -\frac{\vr}{1-\M^2}\Bsh\int_{\rs}^{\rsh}f_0\lam
\e^{\int_{\rsh}^r\frac{i\omp}{\vr}\frac{1+\M^2}{1-\M^2}\d r} \d r \ ,
\label{disp1}
\eeq
where the parameter $\lam$ is:
\beq
\lam \equiv \frac{1}{r^2\cs^2\vr}\left\lbrack\frac{il\omp}
{\vr^2}-im-\frac{2l}{\vr}
\frac{\d\log(r\vr)}{\d r}\right\rbrack \ ,
\eeq
and the value of $\p f/\p r$ at the shock is deduced from Eqs. (\ref{dfdr})
and (\ref{fsh}-\ref{Bsh}).

\section{Reflection at the shock}

Perturbations $\fsh,\gsh$ at the
shock are decomposed on acoustic waves $f_0^{\pm},g_0^{\pm}$ and
advected perturbations $f_{\rm B},g_{\rm
B}$:
\beq
\fsh&=&f_0^++f_0^-+f_{\rm
B} \ ,\label{decompf}\\
\gsh&=&g_0^++g_0^-+g_{\rm
B} \ .\label{decompg}
\eeq
The second order differential equation deduced from the
differential system (\ref{df0}-\ref{dg0}) is
\beq
{\p^2 f_0\over\p r^2}+\left\lbrace {\p\log\over\p
r}\left({1-\M^2\over\M}\right)
-{i\omp\over
\cs}{2\M\over1-\M^2}\right\rbrace{\p f_0\over \p
r}\nonumber\\
+\left\lbrace\omp^2-{m^2\cs^2\over
r^2}+iv_r{\p\omp\over\p
r}\right\rbrace
{f_0\over(1-\M^2)\cs^2}=0 \ .
\eeq
This equation is written in Eq.~(\ref{difcanon}) in a canonical
form as in Kato (1987), with
\beq
k^2&\equiv
&{\omp^2\mu^2\over(1-\M^2)^2\cs^2}
-{1\over2}{\p^2\over\p
r^2}\log{1-\M^2\over\M}\nonumber\\
&&-\left({1\over2}{\p\over\p
r}\log{1-\M^2\over\M}\right)^2 \ .\\
\mu^2&\equiv&1-{m^2\cs^2\over\omp^2r^2}(1-\M^2) \ .
\eeq
In the high frequency limit, the WKB approximations of the ingoing
$f_0^{+}$ and outgoing $f_0^{-}$ acoustic solutions are:
\beq
f_0^{\pm}&\propto&\left({\omega\M\over\omp\mu}\right)^{1\over
2}
\exp\left(\int^r  i\omp{\M\mp\mu\over1-\M^2}{\d r\over
\cs}\right) \ .\label{fpmwkb}
\eeq
The definition of the acoustic flux
used in Foglizzo (2002) can thus be extended to rotating flows as
follows;
\beq
F_\pm\equiv {{\dot M}_0\over
c^2}{\mu\omp\over\M\om}|f_\pm|^2 \ .\label{acousflux}
\eeq
The
variations of $f_0^{\pm}$ are dominated by the phase variations in
the WKB limit:
\beq
{\p f_0^{\pm}\over\p r}&\sim& {i\omp\over
\cs}{\M\mp\mu\over1-\M^2}f_0^{\pm} \ ,\\
g_0^{\pm}&\sim&\pm{\omp\mu\over\om\M}f_0^{\pm} \ .
\eeq
The dominant contribution of the advected vorticity to $f$ and $g$,
at high frequency, is deduced from the differential system
(\ref{dfdr}-\ref{dgdr}) where the derivative were replaced by a
multiplication by $i\omp/v_r$:
\beq
f_{\rm B}&\equiv&
{mv_r^2-l\omp\over\omp^2r^2+m^2v^2}{iB_{\rm sh}\over
\cs^2} \ ,\label{fB}\\
g_{\rm B}&\equiv& {imB_{\rm
sh}\over\omp^2r^2+m^2v^2} \ .\label{gB}
\eeq
Using Eqs.~(\ref{fB},\ref{gB}) and the boundary conditions
(\ref{fsh}-\ref{gsh}-\ref{Bsh}), the linear system
(\ref{decompf}-\ref{decompg}) is solved in order to obtain
$f_0^{\pm}$:
\beq
f_0^{\pm}={\Delta v_r\over
2v_r}{\M\over\mu}{(1-\M^2)(\M\mp\mu)^2\over\pm1-\mu\M}
\left\lbrack1+{i\eta\cs\over\omp
r}{1-\M^2\over\M\mp\mu}\right\rbrack.\nonumber\\
\eeq
The reflection
coefficient ${\cal R}_{\rm sh}$ is defined using
Eq.~(\ref{acousflux}), so that
$|{\cal R}_{\rm sh}|^2$ is the
fraction of the reflected acoustic flux reflected inward:
\beq
{\cal
R}_{\rm sh}&\equiv &{f_0^+\over
f_0^-} \ ,\\
&=&
-
{1+\mu\M\over1-\mu\M}
\left({\M-\mu\over\M+\mu}\right)^2
{1+{i\eta\cs\over\omp
r}{1-\M^2\over\M-\mu}
\over
1+{i\eta\cs\over\omp
r}{1-\M^2\over\M+\mu} \ .
}
\eeq
In the high frequency limit, required
for the validity of Eqs.~(\ref{fpmwkb}) to (\ref{gB}), $\mu\sim 1$
and the reflection coefficient is given by
Eq.~(\ref{rshwkb}).

\section{Approximation of the amplification at
corotation}

Defining a new variable $X$ as in Foglizzo
(2001),
\beq
{\d X\over\d r}\equiv {\M\over1-\M^2} \ ,
\eeq
the
differential equation of acoustic perturbations is simply
\beq
{\p^2
{\tilde f}\over\p X^2}+
\left\lbrack (\Omega-\Omega_0)^2-{\cs^2\over
r^2}(1-\M^2)\right\rbrack
{m^2{\tilde f}\over\M^2\cs^2}=0 \ .
\eeq
This
equation can be approximated as a parabolic cylinder differential
equation in the vicinity of the corotation, by linearizing the
rotation frequency and neglecting the variation of the Mach number in
this region:
\beq
X\sim {\M_0\over 1-\M_0^2}(r-\rco) \ ,\\
{\p^2 {\tilde
f}\over\p r^2}+
\left\lbrack {{\dot\Omega}^2_0\over
\cs^2}{(r-\rco)^2\over1-\M_0^2} -{1\over
\rco^2}\right\rbrack
{m^2{\tilde f}\over1-\M_0^2}=0 \ ,
\eeq
where the
subscript '0' denotes the corotation radius.
The effect of advection
is obtained formally from GN85 with $\kappa=0$ by replacing the sound
speed
$\cs$ by $\cs(1-\M_0^2)^{1\over2}$, and the azimuthal
wavenumber $m$ by $m/(1-\M_0^2)^{1\over2}$.
The fractional amplitude
of the transmitted wave, deduced from Eqs.~(5-8) of GN85 is thus
independent of the Mach number in this approximation:
\beq
{\cal
T}_{\rm co}=\exp\left(-{\pi\over 2} {m\cs\over
|{\dot\Omega}_0|\rco^2}\right) \ .
\eeq
An equivalent way to obtain the
same result is the WKB approximation of the tunelling obtained by
integrating the variation of the radial wavenumber, as in Kato
(1987). Denoting by $r_{\rm IL},r_{\rm OL}$ the two zeros of $k$
deduced from Eq.~(\ref{k2}),
\beq
{\cal T}_{\rm
co}&\equiv&\exp(-\Phi_{\rm LL}) \ ,\\
\Phi_{\rm
LL}&\equiv&-i\int_{r_{\rm IL}}^{r_{\rm OL}}k\d
r \ ,\\
&\sim&\int_{r_{\rm IL}}^{r_{\rm OL}}\left\lbrack {m^2\over
r^2}-
{\omp^2\over(1-\M^2)\cs^2}\right\rbrack^{1\over 2}{\d r\over
(1-\M^2)^{1\over 2}} \ ,\\
&\sim& {\pi\over 2}{m\cs\over
|{\dot\Omega}_0| \rco^2} \ .\label{phiLL}
\eeq

\end{document}